# Two Plateau Moduli for Actin Gels

A.C. Maggs,
ESPCI,
10 Rue Vauquelin,
75231  Paris Cedex 05.

**Abstract:**
Conflicting experimental results have been reported for the plateau modulus in actin solutions: Analogies are often made with the viscoelastic behaviour of flexible polymers making use of the idea of entanglement as the source of the macroscopic storage modulus. We resolve apparent experimental and theoretical contradictions by pointing out the possibility of two distinct plateau regimes as a function of frequency in semidilute solutions of semiflexible polymers. We make the point that longitudinal and transverse hindrance can have very different effects at the macroscopic scale.

**Introduction:**

Purified actin solutions are often taken as a simple mechanical model for the actin cortex of eucaryotic cells. The mechanical properties of the cortex are linked to the mechanical properties of the individual actin filaments, though clearly the properties of the biological system are modified by the presence of a multitude of actin associated proteins [1] which crosslink and regulate the viscoelastic state of the cellular cortex. Understanding of the mechanical nature of the actin cortex must surely pass, however, by an understanding of the mechanical and rheological properties of simplified actin solutions. These solutions are known as gels in the biological literature, even though there is little evidence of crosslinking playing a role in their mechanical properties. Unfortunately there are considerable variations in the reported values of the plateau modulus of purified actin solutions. Several groups have measured the macroscopic viscoelastic properties of actin using various concentration and frequency ranges [2-6]; the values of the viscoelastic constants found vary by many orders of magnitude without any clear trend emerging from the data. The situation has been further confused by the assimilation of results coming from macroscopic rheological experiments and experiments performed with micron sized magnetic particles which, we shall argue, do not measure the same effective elastic constants.

Two recent theoretical papers have approached the problem of the value of the plateau modulus in actin solutions [7,8]. The conclusions were rather different since the mechanisms which were assumed to be at the origin of the plateau modulus were not the same. One article considers the loss of entropy due to variations in tube geometry as the cause of the plateau modulus and finds agreement with low frequency, low density experiments performed by the Munich experimental group [2], while the second paper postulates the presence of entanglements or crosslinking as short length scales to explain the high values of the modulus found by other experimental groups [3] at higher densities and frequencies. Our aim in this paper is to show that one can expect in general two plateau modului as a function of the frequency and that at least part of the variation between the experimental results could be due this fact; however no experimental paper has yet seen the effects that we believe must be present. We wish also to make the point that the rheology of semiflexible polymers could have significant differences with the rheology of flexible polymers and that great care should be taken in the interpretation of experimental results over limited frequency ranges.

We remind readers that flexible semi-dilute polymer solutions have three characteristic time scales in which the viscoelastic behaviour is very different. At the longest times the sample is fluid and is characterised by a macroscopic viscosity. On short time scales one samples the Rouse internal modes leading to a characteristic variation of the viscoelasticity in $\sqrt{\omega}$. Between these two regimes the material has an elastic plateau due to the presence of topological entanglements, or knots which can only relax via the process of reptation. The value of this plateau is determined by the distance between these topological defects[10,11]. The aim of this paper is to determine to what extent similar mechanisms are active in the rheology of semiflexible polymers. We do this in a modified tube model of the dynamics.

Our fundamental hypothesis is that the effect of the surrounding polymer solution behaves as an enclosing tube, as in the dynamics of flexible polymer solutions. However, we consider that the longitudinal dynamics in the tube are unhindered. In this we at variance with the hypothesis in [7] where the effect of surrounding polymers is to block relaxation in both longitudinal and transverse directions. Despite the relaxation of the constraint on the longitudinal motion the solution is still able to store elastic energy due to the long relaxation times of longitudinal modes. Thus we do not use the usual language of entanglements in the description of the dynamics since this implies the presence of steric hindrances which block all relaxation dynamics rather than only transverse dynamics.

Most rheological experiments on actin are carried out in a concentration range where the typical distance between two neighbouring filaments, L, is between 0.3 and 1 micron, a distance small compared with the persistence length $l_p \approx 15\mu$ [9]. The length of the filaments is not very well characterised, the samples are rather polydisperse. We are thus in a novel rheological regime compared to that studied in the classical rheology of flexible polymers where the persistence length is small compared with the distance between contacts or entanglements. However, as in solutions of flexible polymers the long time rheology of actin solutions is dominated by reptation of the filament out of the confining tube. In typical experiments where the mean length of actin filaments is between 20 and 30 microns this time can be many hours, it is the longest time scale in actin rheology experiments. This escape from the tube leads to fluid behaviour as seen in high sensitivity rheological experiments [2]. A detailed discussion of the times scales is given in [7]

In semiflexible polymers there is large discrepancy in time scales between the diffusion of the centre of mass of a filament out of the tube and the slowest of the internal modes of a filament [7]. Over this range of time scales no dissipation can take place and the modulus must show a plateau. The modulus has as its origin the loss of entropy due to distortion of the confining tube; in an equilibrated sample the tube confining a filament has its optimal (lowest free energy) form; shear changes the tube geometry and thus must lead to an increase in the free energy. This modulus quite small in value and can be estimated [7] as $G = k_B T / l_e L^2$ where $l_e$ is the mean distance between collisions of the filament with the tube wall; this modulus is probably smaller than 1 Pascal in typical experimental conditions; $l_e$ has been estimated as $l_e = L^{4/5} l_p^{1/5}$ [12]. L is the mean distance between filaments. Experimentally this modulus has been observed on time scales between 100s and 10000s [2]. L is a function of the actin monomer concentration, c and varies as $L \sim 1/\sqrt{ac}$ with $a$ the monomer size.

Unlike in the rheology of flexible polymers shear couples directly to length fluctuations of filaments; transverse thermal fluctuations in the filament lead to variations in the local density of the filament in the tube and hence an effective longitudinal modulus [13], Fig. 1. Under shear successive sections of the filament of length $l_p$ are then placed under tension or compression depending on their orientation, Fig. 2. Each section of filament of length $l_e$ behaves as a rather hard longitudinal spring [7]. Since the persistence length of actin, $l_p$ is extremely large the initial imposed strains which are coherent over the length $l_p$ must diffuse a large distance in order to begin to relax. This relaxation has a rather well defined characteristic time scale of about 100 seconds and is a source of dissipation and hence variation in the effective modulus. For higher frequencies than this relaxation time the dissipation is again weak and we find a new plateau dominated by the longitudinal elastic constant of the filament. At the shortest time scales (smaller than 0.1 second) the concept of a tube breaks down and one must consider the movement of individual filaments.

In the second part of this article we present a detailed calculation of the dynamics of a filament in a tube. We will conclude that between 0.1 and 100 seconds the sample can be considered as being formed from an ensemble of elementary elastic elements of length $l_e$ with a Hook modulus [7,8] equal to $K_l = l_p^2 k_B T / l_e^4$, leading to a macroscopic shear modulus equal to $K = l_p^2 k_B T / l_e^3 L^2$. The elastic constant has as its origin the crumpling, or ironing out of thermal fluctuations under the influence of the stress acting on a filament. The second value for the modulus is much larger than the first, given above, and can be equal to some tens or hundreds of Pascals. Similar high moduli have been recently proposed but only by supposing the existence of entanglements or crosslinks at the scale $l_e$ [8]. Isambert and Maggs [7] postulated this result as the high frequency limit of the elastic modulus but were unable to deduce the full frequency dependence and the possibility of the second plateau. No entanglement is needed here for producing the high modulus. It exists because there is no relaxation mode available to the system in the time scale between 0.1 and 100 seconds. We give a schematic drawing of the frequency dependence of the modulus in figure 3.

Experiments performed with small magnetic particles in a frequency range near 1 Hz couple directly to the local bending of individual filaments [14,15]. Our explanation for the macroscopic modulus is in terms of coupling to longitudinal density fluctuations in the tube. The effective modulus seen by microscopic particles is much smaller than that measured in macroscopic rheological experiments in the same frequency range. There is no simple way of comparing the results of macroscopic and microscopic rheological experiments. The small values of the modulus found near 1Hz from microrheology can not be interpreted as implying similarly small values for the macroscopic modulus.

We now present a detailed calculation of the motion of a semiflexible polymer in a tube and show how it leads to the above results for the macroscopic rheology of actin filaments. Note that we choose to work in a system of units such as that $k_B T$ and the viscosity of water are taken as unity. In this case time has the units of a volume.

**Reptation of semiflexible polymers:**

We shall deduce the elastic modulus of a solution of actin filament by using the fact that the time dependent modulus $G(t)$ is equal to the stress in a sample after the instantaneous application of a strain $\varepsilon$. $G(t)$ is related to the complex modulus via the equation $G(\omega) = i\omega \int G(t) \exp(i\omega t)\, dt$ [11].

Consider a filament confined in a tube whose mean trajectory in space is given by
$$\mathbf{R}(s) = (x(s), y(s), z(s)) \qquad (1)$$
Where s is the curvilinear distance along the original tube so that
$$|d\mathbf{R}(s)/ds| = 1. \qquad (2)$$
Under an external shear the tube is distorted in space to the form
$$\mathbf{R}(s) = (x(s) + \varepsilon\, y(s), y(s) + \varepsilon\, x(s), z(s)) \qquad (3)$$
Under this distortion we can recalculate the tangent (2) to find
$$|d\mathbf{R}(s)/ds| = 1 + 2\varepsilon (dx/ds)(dy/ds) \qquad (4)$$
The longitudinal modes of the filament in the tube are characterised by an elastic constant, $K_l$, so that the local stress, $T$ in a filament after an initial rapid shear is given by
$$T(s, t=0) \sim K_l\, \varepsilon\, (dx/ds)(dy/ds) \qquad (5)$$
The stress which is due to a local excess or deficit of material in the tube after the shear is subjugated to a local conservation law. If there is a local tension due to a deficit of filament locally it can only be relaxed by the influx of new filament from further along the tube thus:
$$\frac{\partial T(s)}{\partial t} = D \frac{\partial^2 T(s)}{\partial s^2} \qquad (6)$$
With $D$ an effective diffusion constant for relaxation of density fluctuations. One can also consider that this equation is the equation of a series of overdamped springs, giving a normal quadratic dispersion relation. One can estimate $D$ by matching time and length scales at the smallest length scales that such a description remains valid. That is for fluctuations occurring at the scale $l_e$. Such fluctuations relax on a time scale, $\tau_e$ where $\tau_e = l_e^4 / l_p$. Matching this expression to equation (6) one finds that

$$D \sim l_p / l_e^2 \qquad (7)$$

The Green function of Eq (6) is Gaussian thus we can solve for the time evolution of the stress, $T(s,t)$ using standard techniques and find

$$T(s,t) \sim \int T(s',0) \frac{\exp(-(s-s')^2 / 2Dt)}{\sqrt{2Dt}} ds' \qquad (8)$$

Note that $<T(s,t)>$ is zero since a filament can be under either tension or compression. Thus we shall examine the behaviour of $<T^2(s,t)>$ for the case of long filaments where end effects can be neglected. A short calculation gives that

$$<T^2(t)> \sim \varepsilon^2 K_l^2 \int <\dot{x}(s') \dot{y}(s') \dot{x}(s'') \dot{y}(s'')> \frac{\exp(-(s'^2 + s''^2) / 2Dt)}{2Dt} ds' ds'' \qquad (9)$$

where the average on the right hand side is an average over the original tube statistics and we have used the fact that for long filaments one can choose s=0. The original tube distribution is that of a semiflexible polymer in equilibrium. The tangent vector to the filament decorrelates over the correlation length. In a simple approximation we can also take the y and x components of the tangent as independent thus the correlation function in (9) can be written as a product of two tangent correlation functions. Thus we find

$$<\dot{x}(s') \dot{y}(s') \dot{x}(s'') \dot{y}(s'')> = <\dot{x}(s') \dot{x}(s'')><\dot{y}(s') \dot{y}(s'')> \sim \exp(-2|s'-s''|/l_p) \qquad (10)$$

Substituting (10) in (9) one finds the following expression for the time evolution of the stress:

$$<T^2(t)> \sim \varepsilon^2 K_l^2 \int \exp(-2|s'-s''|/l_p) \frac{\exp(-(s'^2 + s''^2) / 2Dt)}{2Dt} ds' ds'' \qquad (11)$$

From this expression one finds two regimes as a function of $Dt / l_p^2$. For short times one can expand the first exponential in (11) in powers of $|s'-s''|$ to find

$$<T^2(t)> \sim K_l^2 \varepsilon^2 (1 - O(\sqrt{Dt} / l_p)) \qquad (12)$$

This is our main result and as announced the stress in the sample remains high until longitudinal stress has had a time to diffuse over the persistence length, $l_p$. which takes $\tau_e = l_e^2 l_p$ or around 100 seconds under the usual experimental conditions. We deduce also that the macroscopic modulus must also remain high over the same time interval and equal to its value at the time $\tau_e$ as stated in the introduction. The macroscopic modulus is found from the spring moduli $K_l$ by noting that a unit volume of material has $L^2$ springs in parallel and $1/l_e$ springs in series implying that $K = K_l l_e / L^2$. Thus K has units of energy per unit volume as expected for a modulus.

For filaments which are long compared with the persistence length one can also consider the regime in which $Dt / l_p^2 >> 1$. In this case the integral is dominated by diagonal band in the $s'$, $s''$ plane (fig. 3). The width of this band, $l_p$ is given by the first exponential in (11) and the length of the band is given by $\sqrt{Dt}$. Thus one can estimate that

$$<T^2(t)> \sim \frac{K_l^2 \varepsilon^2}{2Dt} l_p \sqrt{Dt} \qquad (13)$$

and that the stress decreases in the sample as $t^{-(1/4)}$ after shear. However, can not expect to see this regime in current experiments since the typical filament lengths are comparable to $l_p$, longer filaments are needed to see this effect. In samples with very long filaments the free energy would be dominated by this residual contribution over a certain frequency range (which could be dominate over the first entropic mechanism) leading to a variation in

$$G(\omega) \sim K(l_p l_e^2)^{1/4} \omega^{1/4} \qquad (14)$$

for the complex modulus at low frequencies. We note that the plateau observed in many experiments is not perfect and perhaps shows a certain contribution of the form (14).

The long time tail in (13) is due to the fact that in 1-d diffusion the density at the origin falls very slowly. When the calculation is done with finite filaments the ends can act as sinks of stress and act as adsorbing boundaries in equation (6). One can write the solution of a diffusion equation such as (6) in terms of the eigenvalues and eigenvectors of the operators involved, one finds that

$$G(s,s',t) = \sum_i \varphi_i(s)\varphi_i(s') \operatorname{Exp}(-\lambda_i t) \qquad (15)$$

Thus for the longest times one is dominated by the exponential decay of the slowest mode of eq (15). These adsorbing boundary conditions on the diffusion equation imply that the slow decay in (13) is replaced by an exponential loss of stress in a given filament when $\sqrt{Dt}$ is large compared with the filament length. With a distribution of lengths in the sample one expects that the exact long time decay is rather sensitive to the preparation of the sample, and thus is rather badly reproducible between different samples and even more so between experimental groups. This suggest that systematic studies of the variation seen in elastic properties between samples should be correlated with the length distribution in a sample.

To conclude one expects two plateau regimes. The first at very low frequencies is due to entropic loss after shear. The second plateau comes from the longitudinal dynamics of actin filaments in their tube. Between these two well defined regimes there is a crossover which is strongly dependent on the distribution of filament lengths in the sample.

**Discussion:**

We have shown how the tube dynamics of a semiflexible polymer lead to a number of novel rheological regimes compared with those studied in flexible polymer systems. In particular the linear coupling of length fluctuations to an external shear leads to the existence of a second plateau regime. In this regime there are no topological entanglements, nor is gelling need to produce a relatively high plateau modulus. In actin it is well known that there are a large variety of associated proteins that interact strongly with actin which act as gelling agents. These agents can have reversible or irreversible interactions. It will be of considerable interest to study these agents experimentally and from the theoretical point of view to see how they modify this image of actin dynamics. It would also be useful to make a careful experimental study of the linearity of the response as a function of applied force. Actin samples are extremely fragile and as is well know extremely liable to breakage, however other less drastic effects such as buckling of filaments out of the tube could also lead to novel non-linear effects including entanglement. The present calculations are valid in the limit that filaments are longer than the persistence length. Shorter filaments will loose their stress more rapidly leading to a change in frequency of the crossover between the two regimes.

With long filaments one could expect a phenomenon of work hardening via the following mechanism: Moderate forces on a sample should be large enough to buckle the filament and form hernias similar to those observed in gel electrophoresis of DNA. These hernias ( where the filament buckles out between the posts forming the tube) can be expected to have a major effect on the longitudinal dynamics. Oscillating rheometers might thus be expected to form solutions where real entanglements are present. Study of this kind of phenomena require the simultaneous observation of mechanical properties and filament configuration via fluorescence techniques.

One might hope to have a greater control over the length distribution by using a method of polymerisation of actin in presence of both gelsolin and phalloidin. Gelsolin is an efficient nucleation centre for actin while phalloidin stops depolymerisation of filaments. In this way it should be possible to prepare relatively monodisperse samples. This should permit a systematic study of the crossover between the two regimes proposed in this paper.

**Figure captions:**

Figure 1.

Fluctuation of a filament in a tube formed by neighbouring filaments. The projection of the density onto the average line of the tube leads to a fluctuation in the local density in the tube . The typical distance between collisions of the filament and the tube wall defines the distance $l_e$.

Figure 2.

Schematic representation of the effect of shear on a filament. (A) Shearing a sample results in an effective stretching of object inclined at 45 degrees to the horizontal and compression of objects at 135 degrees to the horizontal. An initial tube (B) is deformed into the form (C). The strains induce stresses which are coherent over a distance equal to the persistence length equal of $15\mu$. The stresses must diffuse over a large distance to begin to relax, leading to a long characteristic time for longitudinal stress relaxation.

Figure 3.

Schematic representation of the plateau modulus, $G(\omega)$ of a solution of actin filaments showing the regimes discussed in the main text: (A) fluid behaviour at very low frequencies. (B) entropic plateau. (C ). crossover between the two plateaux, which is expected to be a sensitive function of the distribution of filament lengths in the solution. (D) plateau due to longitudinal elasticity. (E) crossover to independent filaments.

Figure 4.

Region of the s'. s'' plane which dominates the integral (11) in the limit of long time.

**References:**
[1] P. A. Janmey, *Current Opinion in Cell Biology* **2**, 11-16 (1991).
[2] O. Muller, H. E. Gaub, M. Barmann, E. Sackmann, *Macromolecules* **24**, 3111-3120 (1991). Ruddies R. Goldman W.H., Isenberg G., Sackmann E. *Eur Biophys J* . **22** 309 (1993).
[3] P.A. Janmey, S. Hvidt, J. Lamb, T. P. Stossel, *Nature* **345**, 89-92 (1990).
[4] P. A. Janmey, U. Eutener, P. Traub and M. Schliwa, *J. Cell Biol.* **113**, 155-160 (1991).
[5] D. H. Wachsstock, W. H. Schwarz, T. D. Pollard, *Biophysical J.* **66**, 801-809 (1994).
[6] M. Sato, G. Leimbach, W. H. Schwarz, T. D. Pollard, *J. Biol. Chem.* **260**, 8585-8592 (1985).
[7] H. Isambert, A.C. Maggs Macromolecules **29** 1036-1040 (1996).
[8] F. Mackintosh P. Janmey J. Kas Phys Rev Let. **75** 4425 (1995).
[9] F. Gittes, B. Mickey, J. Nettleton, J. Howard, *J. Cell Biol.* **120**, 923-934 (1993).
[10] P.G. de Gennes *Scaling theory of poymer physics* (Cornell University press)
[11] Doi and Edwards *Dynamics of polymer solutions* (Oxford University press)
[12] Semenov A.N. *J. Chem Soc. Faraday Trans* **2** 317 (1986).
[13] Note that this is a modulus which is generated by the roughening of the filament by thermal fluctuations, the compression modulus of actin is high and does not play a role here.
[14] F. Ziemann, J. Radler, E. Sackmann, *Biophysical J.* **66**, 1-7 (1994).
[15] F. Amblard, A. C. Maggs , B. Yurke, A. Pargellis and S. Leibler Preprint .

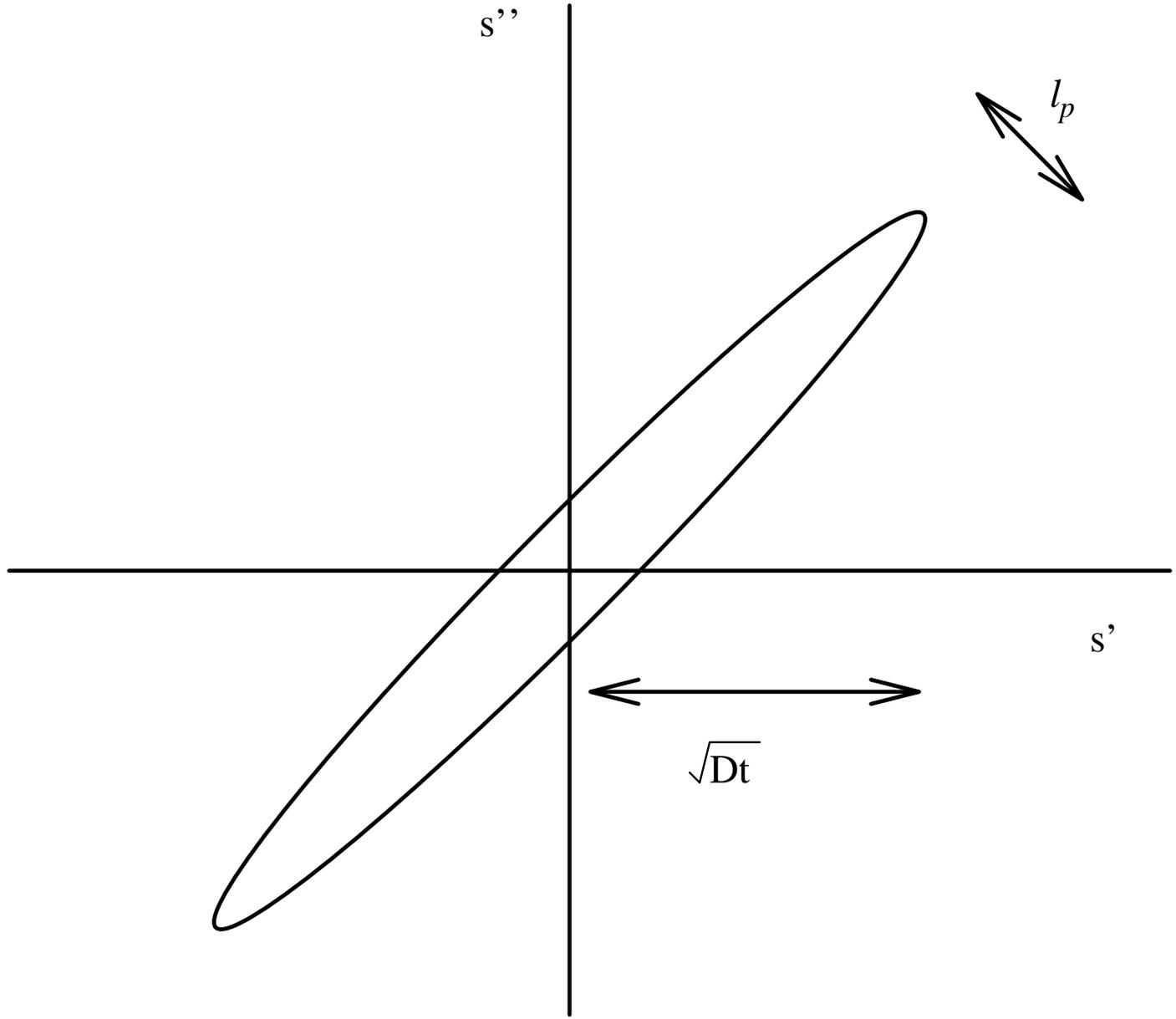

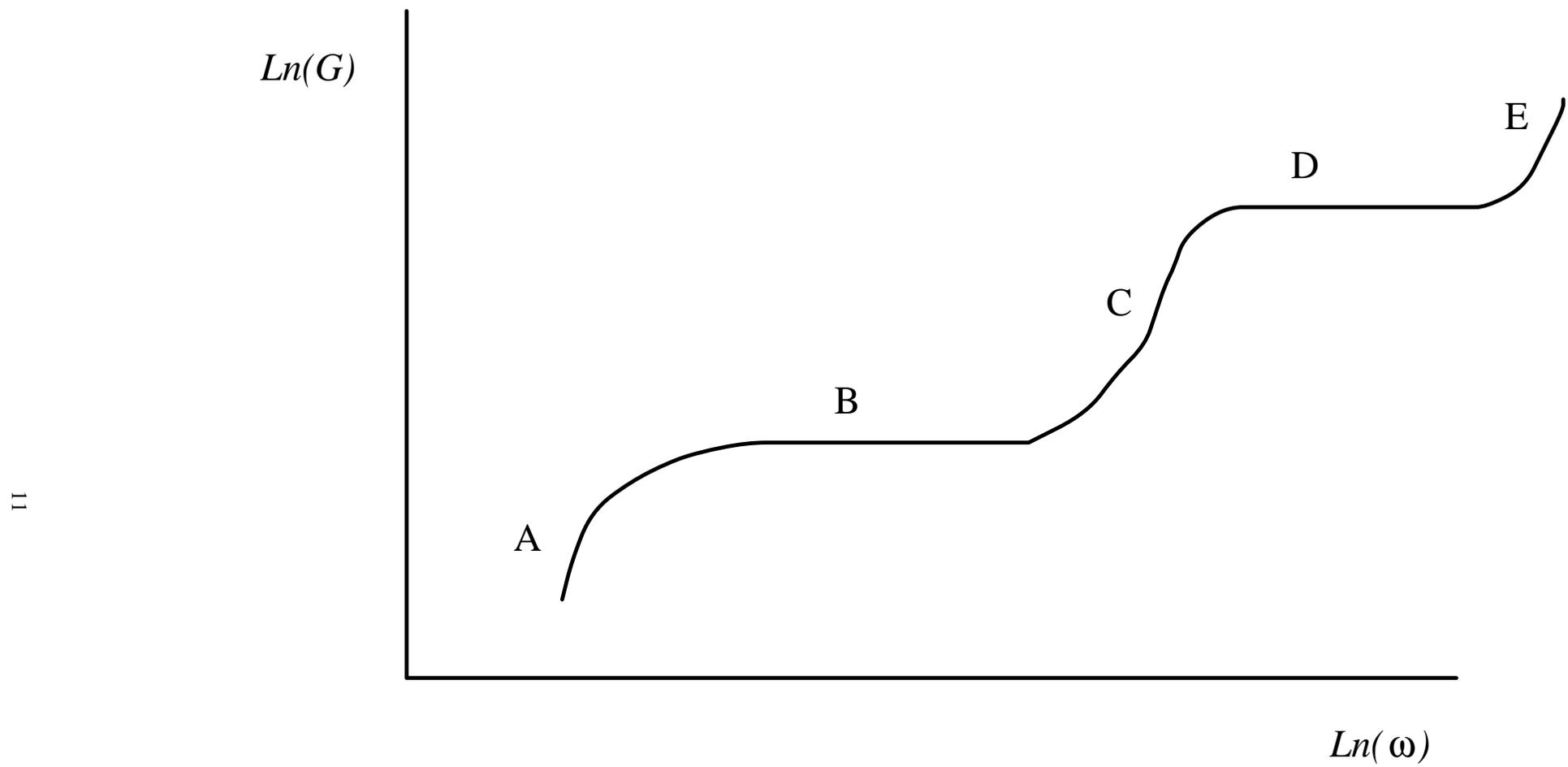

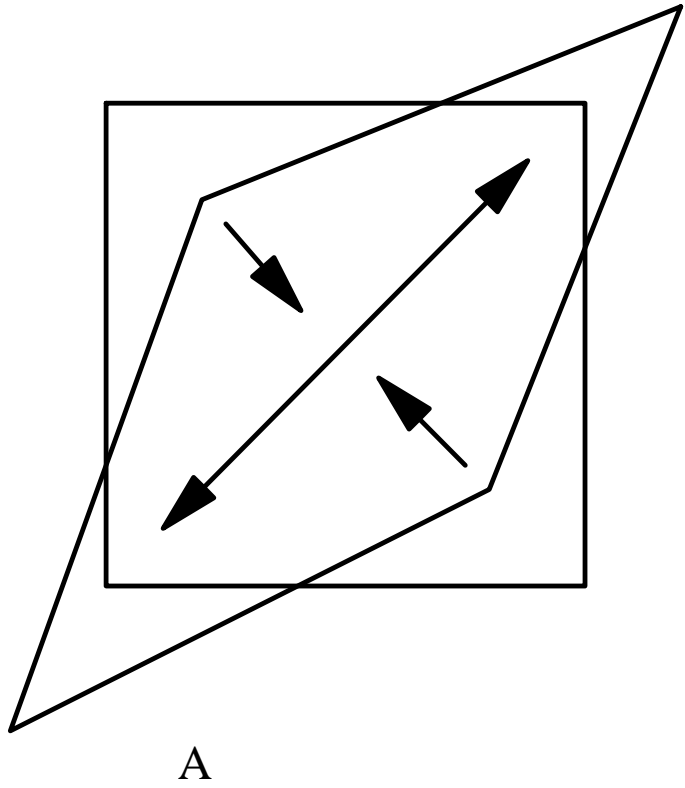
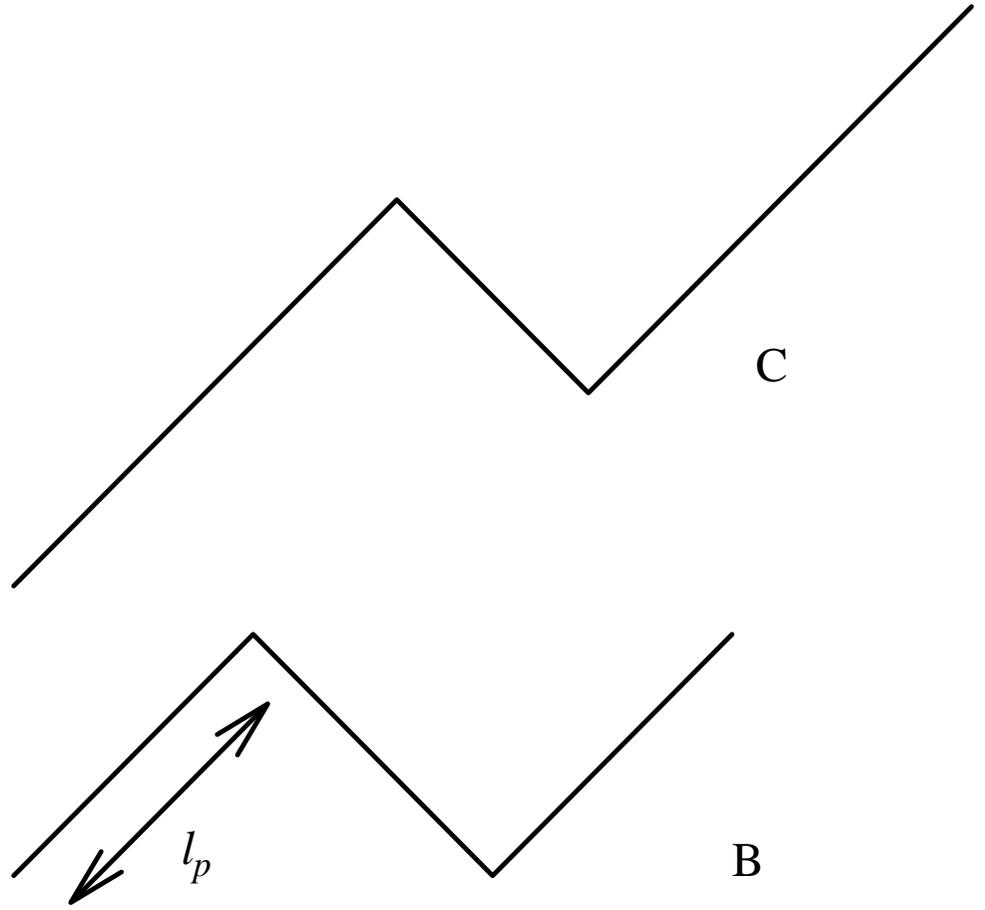

A

B

C

$l_p$

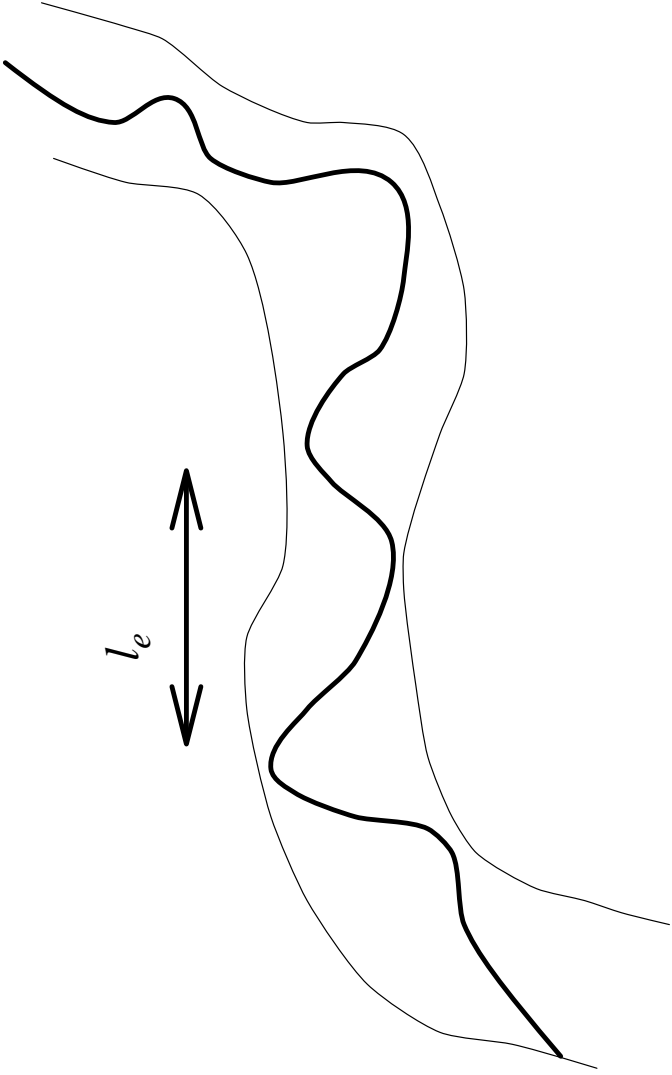